\documentclass[12pt,preprint]{aastex}

\begin{document}

\title{The first photometric analysis of the near contact binary IR Cas}

\author{Li K.\altaffilmark{1,2,3}, Hu, S.-M.\altaffilmark{1,2}, Guo, D.-F.\altaffilmark{1,2}, Jiang, Y.-G.\altaffilmark{1,2}, Gao, D.-Y.\altaffilmark{1,2}, Chen, X.\altaffilmark{1,2}}

\altaffiltext{1}{Institute of Space Sciences, School of Space Science and Physics, Shandong University, Weihai, 264209, China (e-mail: kaili@sdu.edu.cn, likai@ynao.ac.cn (Li, K.), husm@sdu.edu.cn (Hu, S.-M.))}
\altaffiltext{2}{Shandong Provincial Key Laboratory of Optical Astronomy and Solar-Terrestrial Environment, Weihai, 264209, China}
\altaffiltext{3}{Key Laboratory for the Structure and Evolution of Celestial Objects, Chinese Academy of Sciences}

\begin{abstract}
The first photometric analysis of IR Cas was carried out based on the new observed $BVRI$ light curves. The symmetric light curves and nearly flat secondary minimum indicate that very precise photometric results can be determined. We found that IR Cas is a near contact binary with the primary component filling its Roche lobe. An analysis of the $O-C$ diagram based on all available times of light minimum reveals evidence for a periodic change with a semiamplitude of 0.0153 days and a period of 39.7 years superimposed on a secular decrease at a rate of $dp/dt=-1.28(\pm0.09)\times10^{-7}$ d yr$^{-1}$. The most reasonable explanation for the periodic change is the light time-travel effect due to a third body. The period decrease may be caused by mass transfer from the primary component to the secondary. With the decreasing period, IR Cas would eventually evolve into a contact system.
\end{abstract}


\section{Introduction}

A binary star is called a near contact binary if the components are very close to fill their Roche lobes. The typical light curve of near contact binary is an EB-type light variation. The spectral types of their primary and secondary components are A or F and G or K, respectively. Theoretical studies suggest that near contact binaries could be in the intermediate stage between a detached or semi-detached state and a contact state (e.g. Hilditch 2001 and Shaw 1990). Once, a component fills its critical Roche lobe, mass transfer will occur. Therefore, these binaries are important observational targets to study mass transfer in close binaries and understand the evolutionary stages of interacting binaries.

In this context, we chose IR Cas (GSC 03998-02007, V = 11.14 mag) as the target to study. The light curve of IR Cas shows typical EB-type light variation. VanLeeuwen and Milone (1987) observed the $VBI$ light curves of IR Cas using the Rapid Alternate Detection System (RADS), and published differential I light curve. IR Cas displays an apparent variable O'Connell effect during their observation. The orbital period of IR Cas was first analyzed by Zhu et al. (2004), a cyclic oscillation with a period of 53.24 years and an amplitude of about 0.133 days were determined, while it undergoes a secular decrease at a rate of $dp/dt=-1.18\times10^{-7}$ d yr$^{-1}$. No one has analyzed the light curve of IR Cas. Therefore, we chose it as our target to observe.

\section{CCD light curves observation}
IR Cas was observed by using the 1.0-m telescope at Weihai Observatory of Shandong University (WHOT) on October 8th and 9th, November 4th and 7th, 2013. WHOT is a classical Cassegrain telescope with a focal ratio of f/8. A back-illuminated PIXIS 2048B CCD camera from the Princeton Instruments Inc. is mounted to the Cassegrain focus of WHOT. PIXIS 2048B camera with a $2048\times2048$ imaging array, provides a field-of-view (FOV) about $12^\prime\times12^\prime$. The details of the photometry system were described by Hu et al. (2014). Four color light curves of IR Cas were obtained. Typical exposure times are 30 s for B band, 20 s for V band, 10 s for R band and 8 s for I band, respectively. The comparison and check stars are shown in Figure 1, they are near the target and have similar spectral types and visual magnitudes. The observed images were processed by using the Image Reduction (IMRED) and Aperture Photometry (APPHOT) packages in the Image Reduction and Analysis
Facility (IRAF\footnote{IRAF is distributed by the National Optical Astronomy Observatories, which are operated by the Association of Universities for Research in Astronomy under cooperative agreement with the National Science Foundation.}) in the standard fashion . The errors of individual points do not exceed 0.01 mag. The observed four color light curves are displayed in Figure 2. EB-type light curves can be clearly seen. Three new times of light minimum ($2456574.0514\pm0.0001,\, 2456575.0720\pm0.0002$ and $2456604.0014\pm0.0001$)are obtained.

\begin{figure}
\begin{center}
\includegraphics[angle=0,scale=0.5]{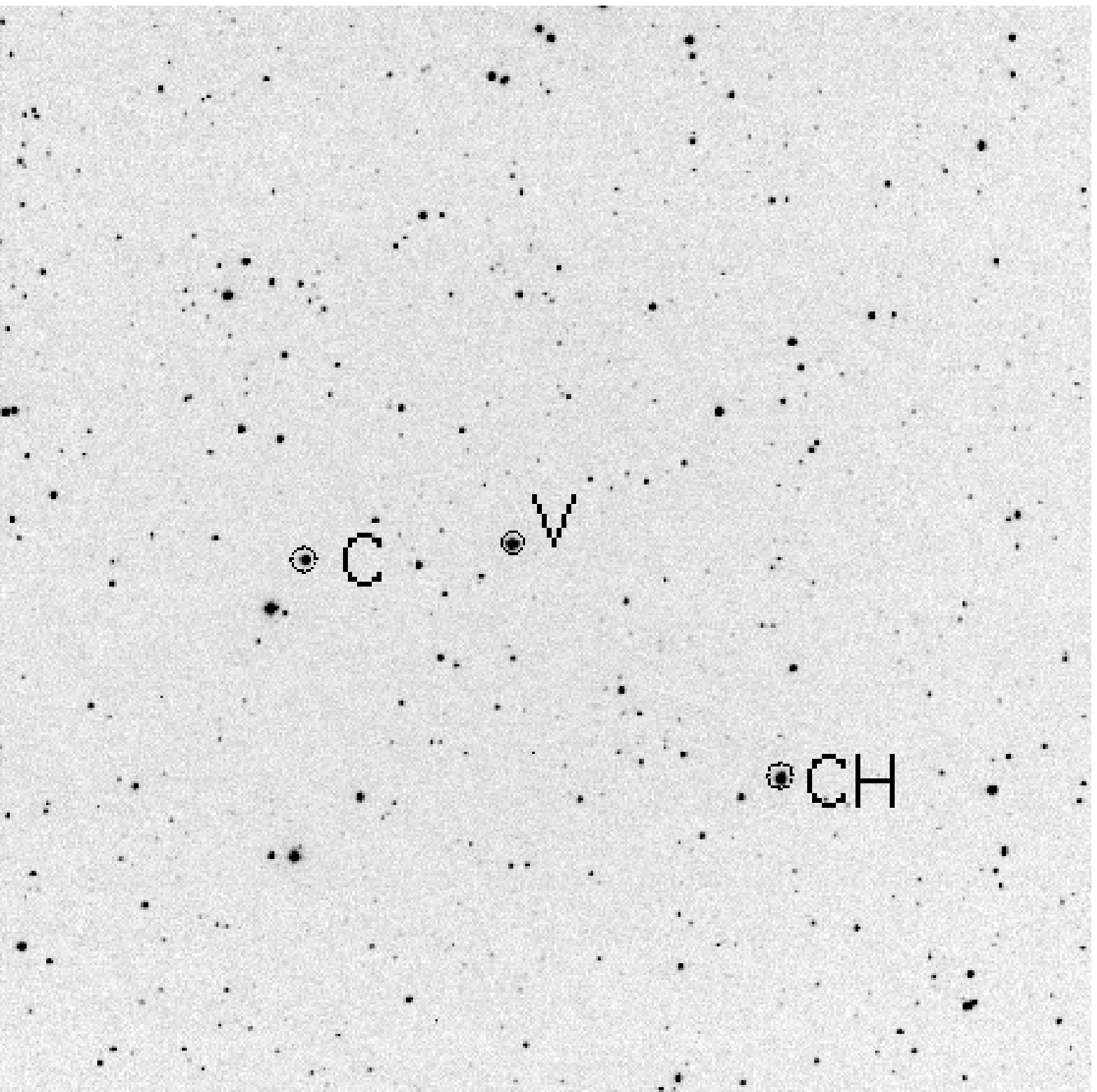}
\caption{The observed image of IR Cas determined by the 1m telescope. V displays the target IR Cas, C represents the comparison star and CH shows the check star. North is up and east is to the left.}
\end{center}
\end{figure}

\begin{figure}
\begin{center}
\includegraphics[angle=0,scale=1.0]{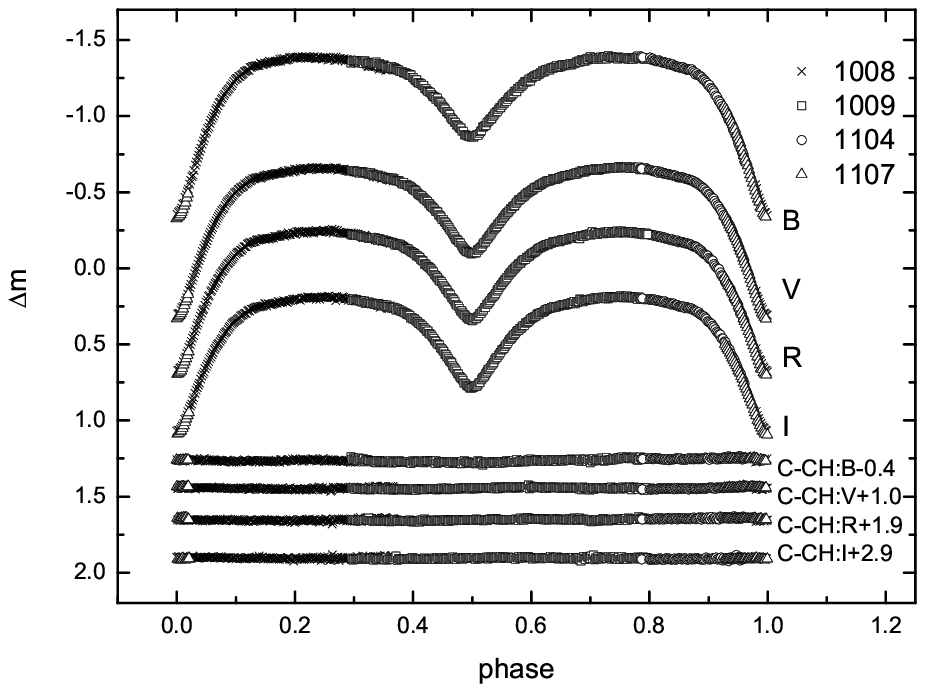}
\caption{The observed four color light curves of IR Cas. Crosses, open squares, circles and triangles represent the light curves observed on October 8th, 9th, November 4th and 7th, respectively. The phases are calculated using the linear ephemeris: Min. I = HJD2456574.0514+0.680686E. }
\end{center}
\end{figure}

\section{The first photometric analysis}

The observed four color light curves of IR Cas were simultaneously analyzed by using the fourth version of the W-D program (Wilson \& Devinney 1971; Wilson 1990, 1994). The symmetric light curves and nearly flat secondary minimum indicate that we can determine very precise photometric results. According to the 4th edition of the GCVS (Kholopov et al. 1987), the spectral type of IR Cas is F4. Therefore, the mean temperature for Star 1 was chosen as $T_1 = 6750$ K. The square root law bolometric and
bandpass limb-darkening coefficients were interpolated from the values of van Hamme (1993): $x_{1bol}=0.108,$ $x_{2bol}=0.167$, $y_{1bol}=0.615,$ $y_{2bol}=0.553$, $x_{1B}=0.224,$ $x_{2B}=0.428$, $y_{1B}=0.658,$ $y_{2B}=0.462$, $x_{1V}=0.083,$ $x_{2V}=0.190$, $y_{1V}=0.710,$ $y_{2V}=0.640$, $x_{1R}=-0.017,$ $x_{2R}=0.075$, $y_{1R}=0.717,$ $y_{2R}=0.667$, $x_{1I}=-0.075,$ $x_{2I}=0.005$, $y_{1I}=0.679,$ and $y_{2I}=0.640$. Following Lucy (1967) and Ruci\'{n}ski (1969), gravity-darkening coefficients of $g_{1,2}=0.32$ and bolometric albedo coefficients of $A_{1,2}=0.5$ were set, which are appropriate for stars having convective envelopes. During our solutions, the adjustable parameters were as follows: the orbital inclination $i$, the mass ratio $q$, the effective temperature of the secondary component $T_2$, the monochromatic luminosity of Star 1 $L_1$ and the dimensionless potential of Star 2 $\Omega_2$.

Since no one has analyzed the light curve of IR Cas, no mass ratio $q$ has been determined, a $q$-search method was used to determine the mass ratio. In order to search the mass ratio $q$, solutions for several assumed values of mass ratio $q$ that is from 0.50 to 1.00 were obtained for IR Cas, solutions for the mass ratio less than 0.50 were also carried out, but the W-D program can not be convergent. For each $q$, the iterations were started at mode 2 (detached mode) and converged at mode 4 (semidetached mode with Star 1 filling the critical Roche lobe). The sums of weighted square deviations $\sum W_i(O-C)_i^2$ for all the assumed values of $q$ are shown in Figure 3. The minimum $\sum$ is determined at $q=0.85$. Then, we chose $q=0.85$ as the initial value and started a differential correction so that the iteration converges by setting $q$ as an adjustable parameter. Finally, the photometric solution is converged at $q=0.851$ and the derived photometric elements are listed in Table 1 (the errors are formal errors only and have no real meaning about the uncertainties). Figure 4 shows the theoretical light curves computed with these photometric elements.

\begin{figure}
\begin{center}
\includegraphics[angle=0,scale=1.0]{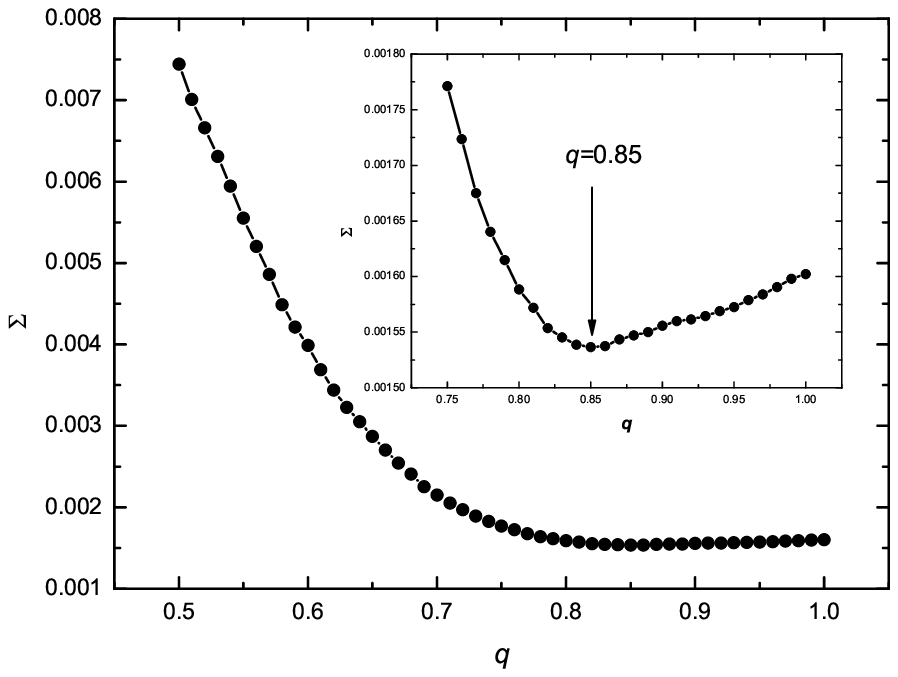}
\caption{$\sum-q$ curves for IR Cas, the small insert figure is an enlargement that $q$ is from 0.75 to 1.00. }
\end{center}
\end{figure}

\begin{figure}
\begin{center}
\includegraphics[angle=0,scale=1.0]{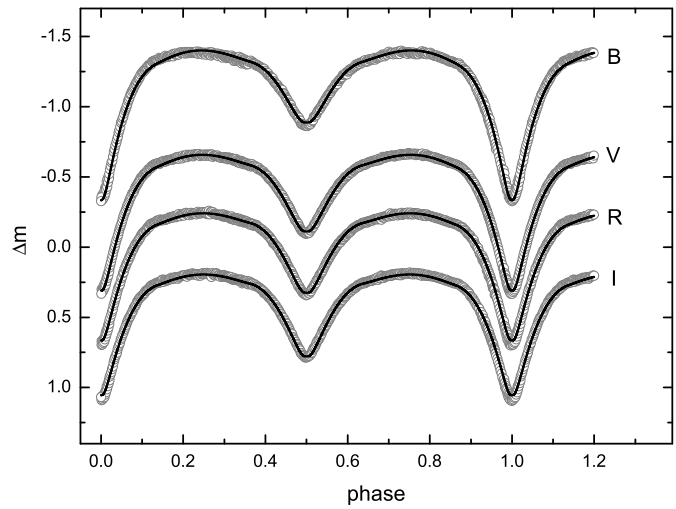}
\caption{Observed and theoretical $BVRI$ light curves of IR Cas. }
\end{center}
\end{figure}

\begin{table}
\begin{center}
\caption{Photometric solutions for IR Cas}
\begin{tabular}{lcc}
\hline
Parameters & Photometric elements & Errors   \\

\hline

  $q(M_2/M_1) $& 0.851  & $ \pm0.005$   \\

  $T_2(K) $&      5992 & $ \pm4$\\

    $i(deg)$ &     86.8 & $ \pm0.2$\\

  $ \Omega_1 $ & 3.5034 & Assumed  \\

  $ \Omega_2 $ & 3.7601& $ \pm0.0145$  \\

$L_{1B}/L_B$&      0.7457 & $ \pm0.0003$\\
$L_{1V}/L_V$&      0.7113 & $ \pm0.0004$\\
$L_{1R}/L_R$&      0.6901 & $ \pm0.0004$\\
$L_{1I}/L_I$&      0.6732 & $ \pm0.0005$\\

  $r_1(pole)$ &   0.3696 &$ \pm 0.0005$\\

  $r_1(side)$ &   0.3891 & $ \pm0.0005$\\

  $r_1(back)$ &     0.4194 & $ \pm0.0005$\\

   $r_1(mean)$ &     0.3932 & $ \pm0.0006$\\

  $r_2(pole)$ &     0.3116 & $ \pm0.0017$\\

  $r_2(point)$ &   0.3601 &$ \pm 0.0037$\\

  $r_2(side)$ &   0.3226 & $ \pm0.0020$\\

  $r_2(back)$ &   0.3417 & $ \pm0.0026$\\

   $r_2(mean)$ &     0.3345 & $ \pm0.0045$\\

\hline
\end{tabular}
\end{center}
\end{table}

\section {Period variation}	
The orbital period variation of IR Cas has been analyzed by Zhu et al. (2004). Based on 313 minima, they found that the orbital period of IR Cas is secular decrease at a rate of $dp/dt=-1.18\times10^{-7}$ d yr$^{-1}$, while it undergoes a cyclic oscillation with a period of 53.24 years and an amplitude of about 0.133 days. Among the 313 minima, only six are photoelectric and CCD minima, the others are visual and photographic minima. Therefore, the results determined by Zhu et al. (2004) are possibly not reliable.

We reanalyzed the orbital period variation of IR Cas by collecting all available visual, photographic, photoelectric and CCD times of light minimum. Using the same linear ephemeris as O-C gateway\footnote{http://var.astro.cz/ocgate/}
\begin{equation}
\textrm{Min.I}=2446332.4510+0.680686\textrm{E},
\end{equation}
we computed the $(O-C)_1$ values for these times of light minimum and listed them in Table 2. The corresponding $(O-C)_1$ curve was displayed in Figure 5. One light minimum time 2441178.3580(BBSAG 31) was discarded because of the very large scatter from other data.
It is found that a secular decrease and a cyclic variation lead to a satisfactory fit to the $(O-C)_1$ curve. Therefore, a second-order polynomial superimposed on a light travel-time effect due to a third body in an elliptical orbit (Irwin 1952), was generally applied to fit the $(O-C)_1$ curve
\begin{eqnarray}
(O-C)_1=T_0+\Delta T_0+(P_0+\Delta P_0)E+{\beta \over 2}E^2+A[(1-e^2){\sin(\nu+\omega)\over(1+e\cos\nu)}+e\sin\omega] \nonumber\\
=T_0+\Delta T_0+(P_0+\Delta P_0)E+{\beta \over 2}E^2+A[\sqrt{(1-e^2 )}\sin E^*\cos\omega+\cos E^*\sin \omega],
\end{eqnarray}
where $T_0$ is the initial epoch and $P_0$ is the orbital period, other parameters are taken from Irwin (1952). In this calculation, different weights were adopted: 8 for the photoelectric and CCD minima and 1 for the visual and photographic minima. The fitted parameters are shown in Table 3. The $(O-C)_2$ values from the quadratic term are plotted in the middle panel of Figure 5. The residuals from the full terms of Equation (2) are displayed in the lower panel of Figure 5. As seen in Figure 5, it is clear that the orbital period of IR Cas varies in a cyclic way, superimposed on the long-term downward parabolic variation. The rate of the secular period decrease is determined as $-1.28(\pm0.09)\times10^{-7}$ d yr$^{-1}$. According to the light travel-time effect term of Equation (2), the period of the cyclic oscillation is about 39.7 years, and the amplitude of which is about 0.0153 days. The results are different from that determined by Zhu et al. (2004).

\begin{figure}
\begin{center}
\includegraphics[angle=0,scale=1.0]{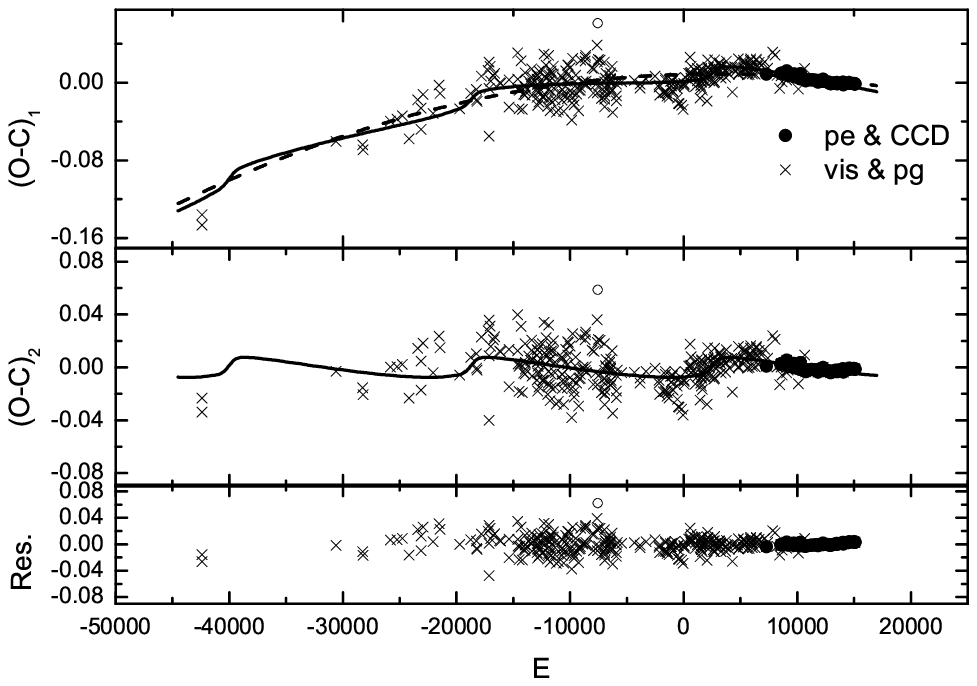}
\caption{$O-C$ diagram of IR Cas. Upper panel displays $(O-C)_1$ curve calculated with the linear ephemeris of Equation (1) based on all available times of light minimum. The $(O-C)_2$ values which remove the secular decrease from the $(O-C)_1$ curve are displayed in the middle. The residuals from full terms of Equation (2) are plotted in the lower panel. Crosses represent visual and photographic data, while filled circles refer to the photoelectric and CCD data. The open circle is a discarded minimum.}
\end{center}
\end{figure}

\begin{table}
\begin{center}
\caption{Times of light minimum for IR Cas}
\begin{tabular}{lccrcccl}\hline\hline

   JD Hel. &         Me &       Type &          E &     $(O-C)_1$ &     $(O-C)_2$ &  Residuals &  Reference \\\hline

2417473.2600  &         pg &          p &     -42397 &   -0.1467  &   -0.0338  &   -0.0267  &  PZ 4, 369 \\

2417473.2707  &         pg &          p &     -42397 &   -0.1360  &   -0.0231  &   -0.0160  & PSMO 16, 244 \\

2425507.4830  &         pg &          p &     -30594 &   -0.0605  &   -0.0030  &   -0.0016  & MHAR 8, 13 \\

2427100.2850  &         pg &          p &     -28254 &   -0.0638  &   -0.0153  &   -0.0114  & MHAR 8, 13 \\

2427120.3600  &         pg &          s &   -28224.5 &   -0.0690  &   -0.0206  &   -0.0167  & MHAR 8, 13 \\

2428750.2920  &         pg &          p &     -25830 &   -0.0396  &    0.0003  &    0.0063  & PSMO 16, 244 \\

2429147.1341  &         pg &          p &     -25247 &   -0.0375  &    0.0004  &    0.0069  & PSMO 16, 244 \\

2429476.5895  &         pg &          p &     -24763 &   -0.0341  &    0.0022  &    0.0090  & PSMO 16, 244 \\
\hline
\end{tabular}
\end{center}
\scriptsize $^1$ This minimum is discarded during the fitting.
(This table is available in its entirety in machine-readable and Virtual Observatory (VO) forms in the online journal. A portion is shown here for guidance regarding its form and content.)
\end{table}

\begin{table}
\begin{center}
\caption{Parameters for the fit of times of light minimum}
\begin{tabular}{lcccc}
\hline\hline

Parameters &     Values &     Errors &    \\\hline

$\Delta T_0$ (d) &     0.0085 &  $\pm0.0014$     \\

$\Delta P_0$ (d) & $3.31\times10^{-7}$ & $\pm0.63\times10^{-7}$ \\

$\beta$ (d/yr) & $-1.28\times10^{-7}$ & $\pm0.09\times10^{-7}$  \\

     $A$ (d) &     0.0153 &  $\pm0.0134$  \\

         $e$ &       0.89 &    $\pm0.22$  \\

   $P_3$ (d) &     14497.3 &   $\pm551.0$ \\

$\omega (^\circ)$ &       10.5 &    $ \pm10.5$  \\

$ T_P$ (HJD) &  2448184.1 &   $\pm317.0 $ \\
\hline\hline
\end{tabular}
\end{center}
\end{table}

\section{Discussion and conclusions}
The first photometric analysis of IR Cas was carried out based on the new observed four color light curves using the 1.0-m telescope at WHOT. The new observed light curves are symmetric, and the secondary minimum is almost flat, indicating that one can determine very precise photometric solutions. It is shown that IR Cas is a semi-detached system with the primary component filling its Roche lobe and can be classified as a near contact binary since the Roche lobe filling ratio of the secondary component is about 93\%. The mass ratio is $q=0.851$ and orbital inclination is $i=86.8^\circ$, the temperature difference between the two components is $\Delta T=758$ K. Since no spectroscopic elements have been published for IR Cas, the absolute parameters cannot be determined directly. Assuming that the primary component is a normal main-sequence star, we can estimate its mass as $M_1=1.43M_\odot$ according to its spectral type of F4 (Cox 2000). Then, the absolute parameters for this system can be determined combining the photometric solutions and its period. They are as follows: $M_2=1.22M_\odot,\, a=4.51R_\odot,\, R_1=1.77R_\odot,\,R_2=1.51R_\odot,\, L_1=5.85L_\odot$ and $L_2=2.63L_\odot$.

Based on all available visual, photographic, photoelectric and CCD times of light minimum, we analyzed the period variation of IR Cas. We found that the secular decrease rate of the period is $-1.28(\pm0.09)\times10^{-7}$ d yr$^{-1}$, the period of the cyclic variation is about 39.7 years. The continuous period decrease means that mass could be transferred from the primary component to the secondary. Assuming a conservative mass transfer and using the Equation
\begin{eqnarray}
{\dot{P}\over P}=-3\dot{M_1}({1\over M_1}-{1\over M_2}),
\end{eqnarray}
we determined that the mass transfer rate is $dM_1/dt=1.13(\pm0.08)\times10^{-6}\,M_\odot$ yr$^{-1}$. The thermal timescale of the primary component can be estimated to be ${GM^2\over RL}\sim7.1\times10^6$ yr, which is very close to the timescale of period decrease $P/(dP/dt)\sim5.3\times10^6$ yr. This may indicate that the mass transfer is on the thermal timescale.

The cyclic period variation could be signs of light time-travel effect due to a third body, or of deformation of a magnetically active companion due to Applegate mechanism (Applegate 1992). Using the relation ${\Delta{P}\over P}=-9{\Delta Q\over Ma^2}$ (Lanza \& Rodon\`{o} 2002), we can calculate the variations
of the quadrupole moment $\Delta Q_{1,2}$ for both components. We determined $\Delta Q_1=1.4\times10^{50}$ g cm$^2$ for the primary component and $\Delta Q_2=1.2\times10^{50}$ g cm$^2$ for the secondary. Assuming conservation of the orbital angular momentum, the typical values of the variation of the quadrupole moment rang from $10^{51}$ to $10^{52}$ g cm$^2$ for active close binaries (Lanza \& Rodon\`{o} 1999). This suggests that Applegate mechanism can not explain the cyclic period variation of IR Cas. Therefore, the cyclic period variation of IR Cas may be more likely caused by the light time-travel effect due to a third body.

The cyclic period and semi-amplitude are calculated to be about 39.7 years and 0.0153 days, respectively. Using the following Equation
\begin{equation}
f(m)={(m_3\sin i^\prime)^3\over (m_1+m_2+m_3)^2}={4\pi\over GP^2_3}\times(a_{12}\sin i^\prime)^3,
\end{equation}
a mass function $f(m) = 0.0118 M_\odot$ is determined for the assumed third body. The masses of the third body for many values of the orbital inclination ($i^\prime$) are computed and are displayed in Figure 6. If we assume that the third body is coplanar to the orbit of the eclipsing pair (i.e., $i^\prime=86.8^\circ$), the values of the mass and the orbital radius of the third body would be
$M_3=0.49\,M_\odot$ and $a_3=17.09$ AU. Assuming that the third body is a main-sequence star, the mass corresponds to the spectral type of M0V. The contribution of the tertiary component to the total light of the system would be less than 1 percent. Therefore, the hypothetical third body would be almost impossible to detect photometrically and spectroscopically with a ground based telescope.

\begin{figure}
\begin{center}
\includegraphics[angle=0,scale=1.0]{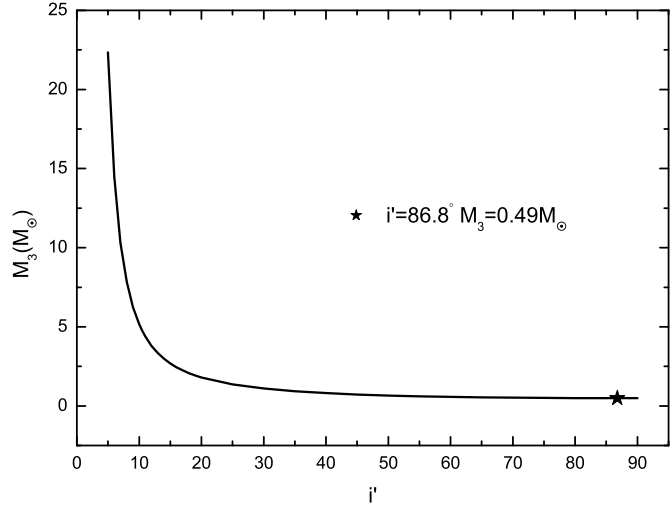}
\caption{Relation between the mass and the orbital inclination for an assumed third body in IR Cas.}
\end{center}
\end{figure}

With the decreasing period, the mass ratio of IR Cas increases and the separation between components decreases. This system will evolve from the present semi-detached phase to contact phase. By considering a mean period of 0.$^d$4 for contact binary stars, it will evolve into a contact binary after $\sim1.2\times10^6$ years. Therefore, it is possible that IR Cas is a progenitor of an evolved contact binary, or is in a broken-contact stage like some interesting systems listed in Table 4. On the other hand, the theory of thermal relaxation oscillations (TRO) models (e.g., Lucy 1976; Flannery 1976; Robertson \& Eggleton 1977; Lucy \& Wilson 1979; Yakut \& Eggleton 2005) predicted oscillation between semi-detached and slightly overcontact configurations. Each oscillation comprises a contact phase followed by a semidetached phase. As we know, these two phases predicted by TRO are quite short compared to the lifetime of the systems. Thus, a system in such phases is very rare. Systems with such configurations are important targets for investigating the evolution of close binaries. IR Cas is a promising candidate for being at this rare evolutionary stage. The photometric solution of IR Cas is only based on photometric observations. For future work, the spectroscopic and photometric observations are needed in order to obtain better understanding of the absolute dimensions and the evolutionary status of IR Cas.

\begin{table}
\begin{center}
\caption{Semi-detached binaries with the primary component filling their inner Roche lobes}
\begin{tabular}{lcccccc}
\hline\hline
Star &     P(d) & Sp.  & $q$ & $dP/dt$(d yr$^{-1}$)  & References\\\hline
BO Peg &   0.58043288 & A7IV  & 0.550 & $-1.26\times10^{-7}$  & (1), (2)\\
BL And &   0.72237705 & A8     & 0.387 & $-2.36\times10^{-8}$  & (3)\\
V473 Cas &   0.4154591 & G2    & 0.493 & $-7.61\times10^{-8}$  & (4)\\
BS Vul &   0.47597147 & G2    & 0.340 & $-2.44\times10^{-8}$  & (5)\\
CN And &   0.46279428 & F5V    & 0.390 & $-1.82\times10^{-7}$  & (6)\\
DM Del &   0.8446747 & G87    & 0.550 & $-2.27\times10^{-7}$  & (7)\\
V388 Cyg &   0.8590372 & A3    & 0.365 & $-2.06\times10^{-7}$  & (8)\\
GO Cyg &   0.717764585 & A0V    & 0.428 & $-1.40\times10^{-9}$  & (9)\\
TT Her &   0.91208023 & F2V    & 0.439 & $-1.82\times10^{-7}$  & (10)\\
IR Cas &   0.680686 & F4    & 0.851 & $-1.18\times10^{-7}$  & (11)\\

\hline
\end{tabular}
\end{center}
\scriptsize
(1)Yamasaki \& Okazaki 1986; (2) Qian 2002; (3)Zhu \& Qian 2006; (4) Zhu et al. 2009; (5) Zhu et al. 2012; (6) Lee \& Lee 2006; (7) He \& Qian 2010; (8) Kang et al. 2001; (9) Ula\c{s} et al. 2012; (10) Milano et al. (1989); (11) This paper.
\end{table}

\acknowledgments
This work is partly supported by the National Natural Science Foundation of China (Nos.
11203016, 11333002, 10778619, 10778701), and by the
Natural Science Foundation of Shandong Province (No. ZR2012AQ008) and by the Open Research Program of Key Laboratory for the Structure and Evolution of Celestial Objects (No. OP201303). We appreciate Ren Dayong for the assistant of the observation of IR Cas. Thanks the anonymous referee very much for her/his very positive comments and helpful suggestions.

\end{document}